\documentclass[twocolumn,trackchanges]{aastex7}

\newcommand{\tcha}{T~Cha}
\newcommand{\Spitzer}{\textit{Spitzer}}
\newcommand{\JWST}{\textit{JWST}}

\newcommand{\NeII}{\textrm{Ne~{\textsc{ii}}}}
\newcommand{\NeIII}{\textrm{Ne~{\textsc{iii}}}}
\newcommand{\ArII}{\textrm{Ar~{\textsc{ii}}}}
\newcommand{\ArIII}{\textrm{Ar~{\textsc{iii}}}}
\shorttitle{Study of PAHs in T Cha using JWST}
\shortauthors{R. Arun}
\submitjournal{AJ}

\begin{document}

\title{When the Wall Fell: Study of Polycyclic Aromatic Hydrocarbons in T Chamaeleontis using \JWST}

\author[orcid=0000-0002-4999-2990,sname='Roy Arun']{R. Arun}
\affiliation{Indian Institute of Astrophysics, Sarjapur Road, Koramangala, Bangalore 560034, India}
\email[show]{arunroyon@gmail.com}  


\begin{abstract}

We investigate the polycyclic aromatic hydrocarbon (PAH) emission features of \tcha, a G8-type T Tauri star that has exhibited ``seesaw''-type mid-infrared continuum variability over nearly two decades due to the destruction of the disk's inner wall using \JWST/MIRI and \Spitzer\ observations. We report the first detection of weak PAH emission at 6.2, 7.7, and 8.6 \micron\ in the \Spitzer/IRS spectrum from 2005. The inner wall destruction in the 2022 \JWST\ epoch allowed more ultraviolet photons to reach the outer disk, increasing the flux levels of PAH bands allowing their detection well above the continuum. The 11.2 \micron\ PAH flux increases by a factor of three, yet its profile shape remains remarkably stable, and the 6.2/11.2 \micron\ \text{flux} ratio has increased, but the charge state of the PAH population remains 75\% neutral. The PAH features exhibit a ``class C'' spectral profile, with redshifted peaks and broadened wings consistent with emission from low-mass T Tauri disks, while the weak 12.7/11.2 ratio points to a lower abundance of duo- and trio-hydrogen modes, implying a predominantly zigzag carbon structure. A faint ``class A'' sub-component in the 6.2 and 7.7 \micron\ bands may indicate additional PAH processing by ultraviolet radiation from accretion hotspots. Placement on PAH charge–size grids locates \tcha\ in the low-ionisation, small-size regime (N\textsubscript{C} $\leq$ 30), signifying a largely neutral PAH population in multiple epochs spanning 18 years. Through multi-epoch, high-resolution data from \JWST\ and \Spitzer, we identify \tcha\ as a benchmark source for probing disk evolution and PAH processing, emphasizing the potential of temporal monitoring with \JWST.

\end{abstract}



\section{Introduction} \label{sec:intro}
Transitional disks, which are characterized by large dust cavities and low near-infrared excess, represent an essential evolutionary stage between optically thick protoplanetary disks and dispersed systems (see \citealp{Espilliat2014prpl.conf..497E}). They have drawn considerable interest because they can exhibit diverse clearing mechanisms, such as photoevaporation, planet–disk interactions, and disk winds \citep{Owen2016PASA...33....5O, Ercolano2017RSOS....470114E}. Over the past two decades, multi-wavelength observations have revealed that some transitional disks show high levels of temporal variability in their infrared continuum \citep{Muzerolle2009ApJ...704L..15M, flaherty2012infrared, espaillat2011spitzer}. This variability can offer unique windows into disk dynamics close to the star, including phenomena such as changes in the inner disk wall or fluctuations in accretion rates. 

\begin{table*}[ht!]
\centering
\caption{Observation log for mid-IR spectra of \tcha\ used in the study.}
\label{tab:obs_log}
\begin{tabular}{c c c c}
\hline \hline
\textbf{Mission/Instrument} & \textbf{Observation Date} & \textbf{Proposal ID/AOR} & \textbf{PI} \\
\hline
\textit{Spitzer}/IRS        & 2004 Jul 18              & 5641216                  & N. Evans     \\
\textit{Spitzer}/IRS        & 2005 May 30              & 12679424                    & J. R. Houck  \\
VLT/ISAAC (L band)          & 2006 Apr 18              & --                   & G. Geers     \\
\textit{JWST}/MIRI MRS      & 2022 Aug 13--14          & 2260                 & I. Pascucci  \\
\hline
\end{tabular}
\end{table*}

Within this framework, \tcha\ has emerged as an intriguing and well-studied example. \tcha\ is a young (2–10 Myr; \citealp{fernandez2008kinematic}), accreting ($\sim6 \times 10^{-9} M_{\odot},\mathrm{yr}^{-1}$; \citealp{Cahill2019MNRAS.484.4315C}) G-type star surrounded by a disk containing a substantial inner gap (from $\sim$0.2 to 10 au), which has been imaged through near-infrared interferometry and millimeter observations \citep{olofsson2013sculpting, huelamo2015high, pohl2017new}. The system exhibits large-amplitude optical variability, including episodic outbursts, and has displayed a pronounced ``seesaw" signature, which refers to an inverse change in the mid-infrared (mid-IR) continuum at short and long wavelengths, on multi-year timescales \citep{schisano2009variability, Cahill2019MNRAS.484.4315C}.

Using the James Webb Space Telescope (\JWST)\ and archival \Spitzer\ data, combined with optical and infrared photometric variability, \cite{Chengyan2025ApJ...978...34X} studied this seesaw effect in \tcha. Their study shows that between the earlier \Spitzer\ data and more recent \JWST/Mid-Infrared Instrument (MIRI) observations, the flux at shorter mid-IR wavelengths (roughly 5–10 \micron) decreased, whereas the flux at longer mid-IR wavelengths (beyond about 15 \micron) increased. This flip in flux, where one wavelength region goes down while another goes up, resembles the two ends of a seesaw moving in opposite directions \citep{Muzerolle2009ApJ...704L..15M,Espaillat2024ApJ...973L..16E}. In \tcha, \cite{Chengyan2025ApJ...978...34X} interpret this behavior as a result of changes in the geometry of the inner disk: as the inner dust wall becomes less tall or less massive, more radiation reaches the outer or shadowed disk, enhancing longer-wavelength emission. Conversely, as the inner wall is destroyed, near- and mid-IR emission from the innermost dust decreases, leading to a dip in the short-wavelength mid-IR range but a rise in the longer-wavelength mid-IR.

Earlier mid-IR spectra acquired with \Spitzer\ suggested the presence of 11.2 \micron\ polycyclic aromatic hydrocarbons (PAHs) \citep{Geers2006A&A...459..545G}, indicating that ultraviolet (UV) photons from the star penetrate regions of the disk’s surface layer. More recently, \cite{Bajaj2024AJ....167..127B} reported that the \JWST\ spectra show 6.2, 7.7, and 8.6 \micron\ PAH features, which were not previously reported in the older \Spitzer\ data. PAHs serve as tracers of UV radiation and are sensitive to the physical conditions in the disk’s upper layers \citep{tielens201125, Geers2006A&A...459..545G}. Their emission profiles are often grouped into classification schemes, commonly referred to as classes A, B, C, and D—based on band peak positions and shapes \citep{Peeters2002A&A...390.1089P,van_Diedenhoven2004ApJ...611..928V, Matsuura2014MNRAS.439.1472M}.

UV photons excite PAH molecule, which then relaxes through various vibrational modes and produces the infrared features that shows up in the astrophysical spectra. The near-IR PAH feature at the 3.3 \micron\ region is due to the C–H stretching mode. The 6.2 and 7.7 \micron\ band features are attributed to C–C stretch modes\citep{Tielens2008ARA&A..46..289T}, and they shift systematically to the redder wavelengths in different environments. In Class A sources (H{\sc ii} regions, reflection nebulae) they peak at 6.22 – 6.25 and 7.6 \micron. Class B spectra (planetary nebulae, Herbig Ae/Be stars) move to 6.25–6.28 and 7.8 \micron, whereas Class C spectra from cool T Tauri disks peaks at 6.28–6.30 \micron\ and exceed 8 \micron. The 8.6 \micron\ line traces C–H in-plane bending modes \citep{Peeters2002A&A...390.1089P,van_Diedenhoven2004ApJ...611..928V}. The 11.2 \micron\ band arises from solo C–H out-of-plane bends in large, neutral PAHs; its weaker 11.0 \micron\ companion comes from cations, so their ratio traces ionization \citep{Peeters2002A&A...390.1089P}. The 12.7 \micron\ band stems from trio (and minor duo) C–H out-of-plane bends \citep{Tielens2008ARA&A..46..289T}.

In this paper, we use the \JWST/MIRI Medium Resolution Spectrometer (MRS) and \Spitzer\ spectra of \tcha\ to study the continuum and PAH variability and characterize the newly identified PAH bands. In the section 2, we describe the observational details of both \Spitzer, \JWST\, and VLT/ISSAC data sets. The section 3, describes the analysis and results, followed by a brief discussion of UV hardness and PAH charge and size in Section 4.  We conclude with a summary of the work with a future scope in Section 5. 

\section{Observations and Data Reduction} \label{sec:observations}

The study is based on three archival mid-IR spectra from \Spitzer, VLT/ISAAC, and \JWST\ spanning nearly two decades. The first dataset is of archival \Spitzer\ InfraRed Spectrograph (IRS) short high (SH) and long high (LH) observations from 2004 \citep{Geers2006A&A...459..545G}, which provided the earliest high-quality view of \tcha’s mid-IR emission. The second data consist of short low (SL) and long low (LL) observation along with \texttt{SH+LH} spectra observed in 2005 \citep{Brown2007ApJ...664L.107B}. The \texttt{SH+LH} spectra in 2005 has a noisy spectra, thus we use \texttt{SL} mode for the work. We retrieved these data from the enhanced offline version of Combined Atlas of Sources with \Spitzer\ IRS Spectra (CASSIS\footnote{CASSIS is a product of the IRS instrument team, supported by NASA and JPL.}) archive \citep{CASSIS2011ApJS..196....8L,Lebouteiller2015ApJS..218...21L}, CASSISjuice \citep{CASSISJUICE2023arXiv230906876L}. The spectral coverage included both the SH and LH modules, extending roughly from 10 to 35 \micron\ at a resolving power of \(R \approx 600\) in the high-resolution configuration or \(R \approx 60\) in low-resolution mode, depending on the particular observation. \autoref{tab:obs_log} shows the details of observation.
\begin{figure*}[ht!]
\centering
\includegraphics[width=2.1\columnwidth]{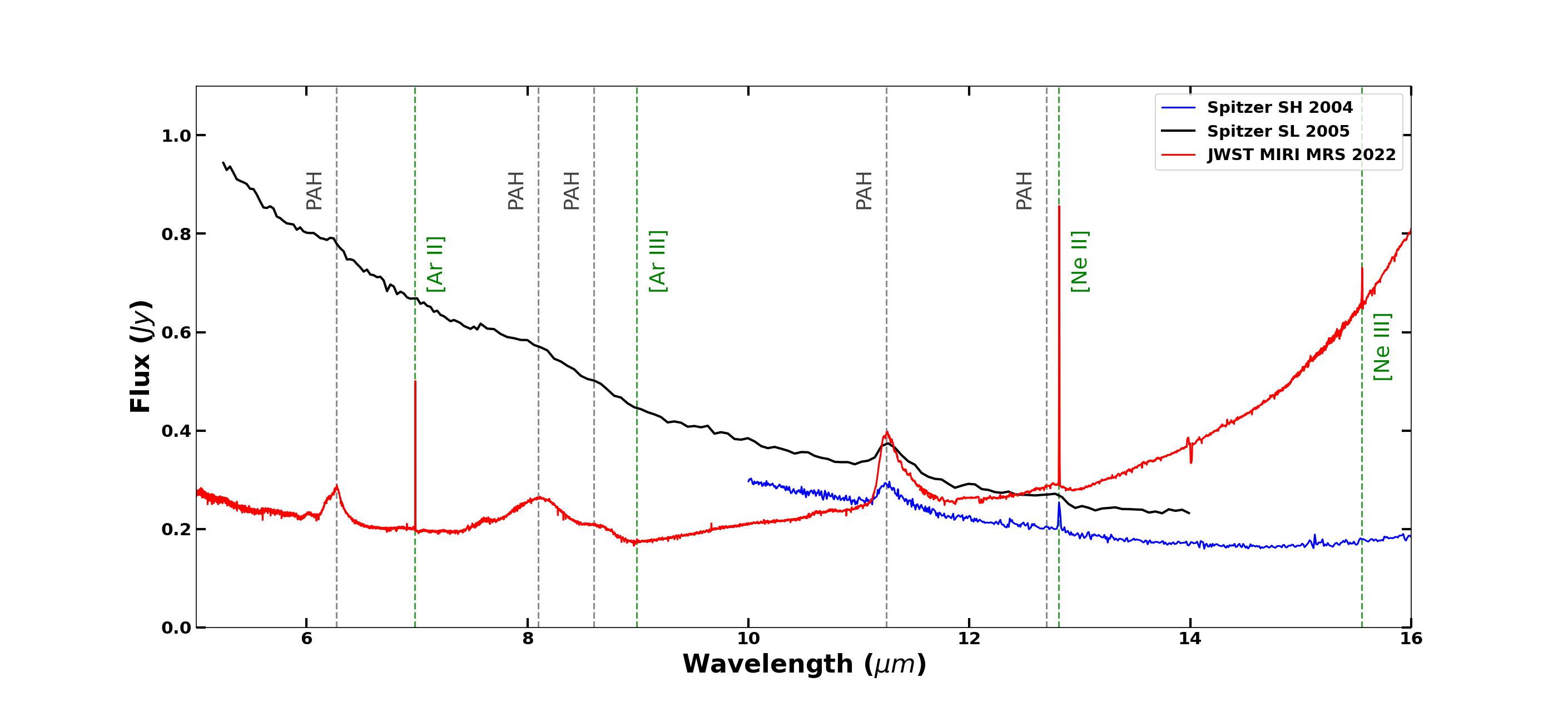}
\caption{
Comparison of the \textit{JWST}/MIRI MRS spectrum of T~Cha (red) with two \textit{Spitzer}/IRS datasets: the 2004 SH module (blue) and the 2005 SL module (black). The 2005 data covers the 5--15 \micron\ range, while the 2004 SH observation starts at 10 \micron. The ``seesaw'' effect is evident, with the \textit{JWST} continuum showing diminished flux below 10 \micron\ but enhanced emission at longer wavelengths, compared to the 2005 \textit{Spitzer} spectra. Forbidden lines ([Ar II], [Ar III], [Ne II], [Ne III]) and the main PAH bands are marked with vertical dashed lines.}
\label{fig:Spitzer_JWST_comparison}
\end{figure*}

The third dataset is a newly acquired \JWST/MIRI MRS observation obtained on 2022 August 13–14 as part of the Cycle 1 General Observer program (Proposal ID: 2260; PI: I. Pascucci). As reported in \citet{Bajaj2024AJ....167..127B} and \citet{Sellek2024AJ....167..223S}, these observations totaled 3.24 hr on-source (4.48 hr including overheads) and covered the full range of the MIRI MRS from approximately 5 to 28 \micron. The MRS operates in four channels, each divided into three sub-bands, yielding 12 individual spectral segments which were subsequently stitched to form a single continuous spectrum at a resolving power of R$\sim$3000. A detailed reduction step can be found in \cite{Bajaj2024AJ....167..127B}. The calibrated \JWST/MIRI/MRS spectra offer a significantly improved sensitivity and spectral resolution over the \Spitzer\ IRS data, enabling a comparative study of PAH variability.

The VLT/ISAAC L-band observations of \tcha\ on 2006 April 18 captured the 3.3 \micron\ PAH feature \citep{Geers2007A&A...476..279G}, which is not covered in the \JWST/MIRI and \Spitzer\ wavelengths. Owing to the lack of \JWST/NIRSpec data, this 3.3 \micron\ feature is used in the study complementing the mid-IR PAH analysis from \JWST.

\section{Analysis and Results} \label{sec:results}
\subsection{Understanding the Variability in the Continuum Emission}

To investigate the continuum evolution of \tcha, we compare the \JWST/MIRI MRS spectrum (2022) with two archival \Spitzer/IRS datasets obtained in 2004 and 2005. The \Spitzer\ \texttt{SH+LH} data from  2004 are mentioned in \citet{Chengyan2025ApJ...978...34X}, but the focus is on the \Spitzer\ \texttt{SL} spectrum from 2005, as the 2004 \texttt{SH+LH} dataset is not directly comparable over the 5–15 \micron\ range. We include the 2004 dataset as an additional data point, as we are looking at the variability of PAHs in different epochs. 

\autoref{fig:Spitzer_JWST_comparison} compares the \JWST spectrum of \tcha\ (red line) with the archival \textit{Spitzer}/IRS data taken in 2004 (blue line; SH module) and 2005 (black line; SL module). The 2004 SH data provide coverage at wavelengths above 10 \micron, but it clearly shows the 11.2 \micron\ PAH emission. Several noteworthy differences are, the flux level in the 5–10 \micron\ range is substantially lower in the \textit{JWST} spectrum than in either \textit{Spitzer} dataset, while at wavelengths longer than about 12–13 \micron\ the \textit{JWST} continuum rises relative to earlier epochs. This behavior exemplifies the “seesaw” effect reported by \cite{Chengyan2025ApJ...978...34X}, wherein the short-wavelength mid-IR flux decreases as the longer-wavelength emission increases. The 2004 and 2005 \textit{Spitzer}\ datasets shows a similar slope in continuum suggestive of similar disk structure. The higher flux in 2005 data over 2004 might be due to with the larger aperture of \textit{Spitzer} low resolution data. 

In the \JWST/MIRI spectrum, all prominent PAH features previously not reported in the older \Spitzer\ data are clearly observed, with the exception of the well-known 11.2~\micron\ feature, which was already reported in earlier observations \citep{Geers2006A&A...459..545G}. Specifically, \citet{Bajaj2024AJ....167..127B} report PAH bands at 6.02 \micron\ and 8.22 \micron, as well as ubiquitous features at 6.2, 7.7, 8.6, 12.0, and 12.7 \micron\ in the \JWST\ dataset, which were not reported by previous \Spitzer\ studies. The study demonstrated that PAH emission is spatially extended, confirming it primarily traces the disk's illuminated surface layers. The clear detection of all PAH bands in the \JWST\ epoch might be due to the diminished inner dust wall that permits more UV flux to reach the PAH-rich surface layers in the outer disk allowing conditions for increased excitation of PAHs \citep{Chengyan2025ApJ...978...34X,Bajaj2024AJ....167..127B}. The higher sensitivity of \JWST\ aided as it raised the signal-to-noise of the PAH features.

The wall destruction and subsequent illumination of outer disk gives us a unique opportunity to study the PAH population in a transitional disk. In the following sections, we provide an comprehensive analysis of individual PAH bands, studying their variability in different epochs and  comparing their properties with some reference young sources.

\subsection{Processing of the Spectra}
For further analysis and comparison, we process the all multi-epoch spectra of \tcha. The process of continuum subtraction and flux estimation of all spectra are described below.

\subsubsection{Continuum Subtraction of the Spectra}

To isolate the PAH emission features and for robust flux measurements across multiple epochs, we performed continuum subtraction using a cubic spline interpolation technique.

For the \JWST\ and \Spitzer/IRS SL spectra (5–15\,\micron\ region), we adopted continuum anchor points from \citet{Seok2017ApJ...835..291S} and \citet{Arun2023MNRAS.523.1601A}, optimized for PAH-dominated young stars. These include nodes at 5.55, 5.80, 6.70, 7.00, 9.15, 9.45, 9.70, 10.20, 10.70, 11.80, 12.20, 13.05, 13.25, 13.80, 14.00, and 14.80 \micron.

For the higher-resolution SH spectrum (10–15 \micron\ region), we used anchor points at 10.20, 10.7, 11.80, 12.20, 13.05, 14.00, and 14.99 \micron. These intervals were selected to avoid known PAH bands and emission lines while preserving the integrity of the underlying continuum shape.

For the VLT/ISAAC spectra, we applied a continuum subtraction described by \citet{Geers2007A&A...476..279G}. The continuum was modeled as a linear function fitted over selected anchor regions free of emission features. The anchor points were chosen around the target features, following the continuum intervals used in PAH studies (e.g., at 3.05--3.07, 3.10--3.70, 3.45--3.50, 3.667--3.689, and 3.7068--3.720 \micron). Once the linear baseline was established, it was subtracted from the observed spectrum.

\subsubsection{Gaussian Decomposition of the 6 \micron\ PAH Complex} \label{subsec:6umdecomp}
 
\begin{figure}[]
\centering
\includegraphics[width=\columnwidth]{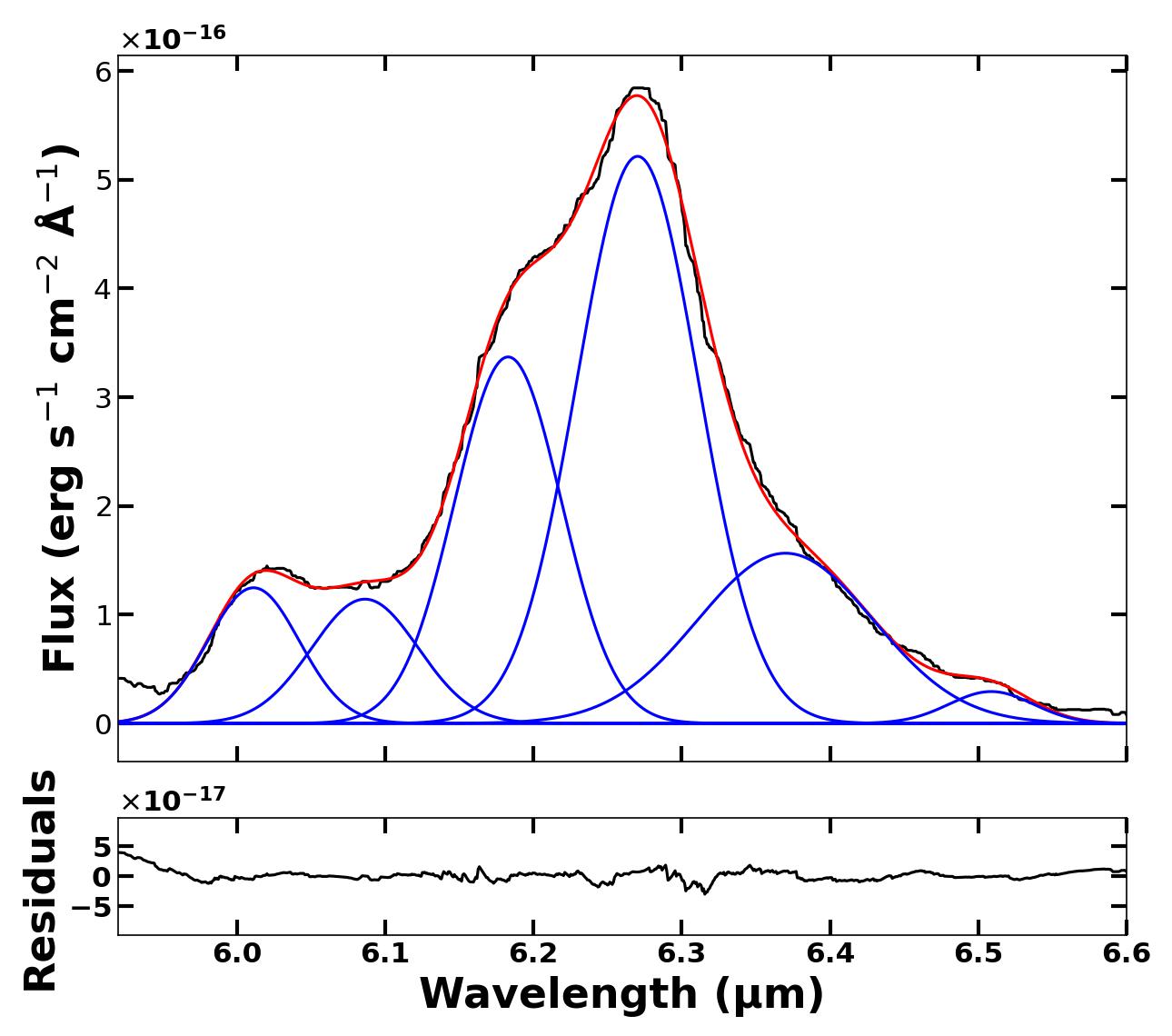}
\caption{Gaussian decomposition of the 6 \micron\ PAH emission in \tcha. The median-filtered spectrum is shown in black. The coloured lines indicate individual gaussian components, and the red line marks the total fitted model. The lower panel plots residuals (data minus model). The primary emission peak at 6.273 \micron\ is significantly shifted relative to the more commonly reported 6.2 \micron\ feature, characteristic of border class B/class C PAH spectra.
}
\label{fig:6_micron}
\end{figure}

The analysis of PAH emission features is often enhanced by decomposing the spectral profiles into multiple Gaussian components. Following the methodology of \citet{ref:Peeters_2024}, who modeled the 3 \micron\ PAH region using ten gaussians (nine distinct PAH bands and one underlying plateau), we similarly performed a multi-component Gaussian decomposition of the 6 \micron\ PAH region. This region was fit with six gaussian profiles (C1-C6). During the fitting process, the central wavelengths of each gaussian component were constrained to lie within ±0.025 \micron\ of their nominal positions, and the FWHM values were restricted to ±0.0025 \micron. This decomposition allows for the identification and separation of the 6.0 \micron\ feature potentially associated with C=O (carbonyl) stretching vibrations \citep{Peeters2002A&A...390.1089P} or olefinic C=C bonds \citep{Hsia2016ApJ...832..213H} and also from other PAH sub components. To examine the detailed substructure of the 6 \micron\ PAH complex in \tcha, we first performed median filtering on the continuum-subtracted spectrum to mitigate narrow emission spikes and instrumental artifacts. Specifically, we employed a median filter with a kernel size of 31 data points, which effectively suppresses sharp features while preserving the broad PAH emission. We then restricted our analysis to the wavelength range 5.95--6.55 \micron, where the strongest PAH emission is expected in this system (see \autoref{fig:6_micron}).

The gaussian components, whose initial parameters are listed in \autoref{tab:6um_init_params}.
\begin{table}[ht!]
\small  
\centering
\caption{Parameters and integrated fluxes from the multi-gaussian fit of the 6 \micron\ region in T~Cha, rounded to 3 significant figures.}
\label{tab:6um_components}
\begin{tabular}{l@{\hskip 6pt}c@{\hskip 6pt}c@{\hskip 6pt}c@{\hskip 6pt}c}
\hline\hline
Component & Amplitude & Mean & $\sigma$ & Flux \\
          & ($\mathrm{erg\,s^{-1}\,cm^{-2}\,\text{\AA}^{-1}}$) & ($\mu$m) & ($\mu$m) & ($\mathrm{erg\,s^{-1}\,cm^{-2}}$) \\
\hline
C1 & 1.25$\times10^{-16}$ & 6.01 & 0.031 & 9.68$\times10^{-14}$ \\
C2 & 1.14$\times10^{-16}$ & 6.09 & 0.036 & 1.03$\times10^{-13}$ \\
C3 & 3.37$\times10^{-16}$ & 6.18 & 0.036 & 3.04$\times10^{-13}$ \\
C4 & 5.21$\times10^{-16}$ & 6.27 & 0.040 & 5.34$\times10^{-13}$ \\
C5 & 1.57$\times10^{-16}$ & 6.37 & 0.060 & 2.33$\times10^{-13}$ \\
C6 & 2.93$\times10^{-17}$ & 6.50 & 0.030 & 2.08$\times10^{-14}$ \\
\hline
\multicolumn{4}{r}{\bf Total Flux} & \bf 1.29$\times10^{-12}$ \\
\hline
\end{tabular}
\end{table}

The multi-gaussian fitting is performed using Levenberg--Marquardt minimization (LevMarLSQFitter) from \textit{astropy} \citep{astropy:2022} to converge on the best-fit parameters for each gaussian. We then numerically integrated each fitted component over wavelength to obtain its flux. \autoref{fig:6_micron} shows the median-filtered data (black line), individual gaussian components (blue), their sum (red line), and the residuals. In general, the fit reproduces the broad substructures of the 6 \micron\ PAH complex well.

\autoref{tab:6um_components} lists the integrated \text{flux} of each fitted gaussian component and the total PAH flux for \tcha’s 5.95--6.55 \micron\ region. We find that the strongest emission peaks near $\lambda\approx6.27$ \micron, consistent with border class B/class C type PAHs, which are frequently shifted to longer wavelengths compared to the class A, 6.2 \micron\ band \citep[e.g.][]{Peeters2002A&A...390.1089P}. The combined flux from the 6 \micron\ region is ${\sim}1.29\times10^{-12}$ $\mathrm{erg\,s^{-1}\,cm^{-2}}$. 

\subsubsection{Estimation of PAH Fluxes}

We estimated the integrated fluxes of the 6.2, 7.7, 8.6, 11.2, and 12.7 \micron\ PAH bands in the \JWST\ spectrum of \tcha\ using direct numerical integration of the continuum-subtracted spectrum. The total flux of 6.2 \micron\ PAH feature (sum of all six components) matches with the integrated flux. The PAH features in \tcha\ are shifted due to their class C characteristics (discussed in following sections), integration intervals are chosen accordingly to encompass the redward extension and asymmetry typical of class C profiles. The 6.2 \micron\ PAHs are integrated between 5.9-6.6 \micron, similar to the gaussian decomposition range. The 7.7 \micron\ complex was integrated between 7.0–8.45 \micron\ and the 8.6 \micron\ band over 8.45–9.05 \micron. For the 11–13 \micron\ region, we adopted 10.9–11.13 \micron\ for the 11.0 \micron\ band, 11.13–11.75 \micron\ for the 11.2 \micron\ band, and 12.2–13.1 \micron\ for the 12.7 \micron\ feature. The final flux of 12.7 \micron\ is estimated after the subtraction of [Ne II] 12.81 line flux. The resulting fluxes are tabulated in the \autoref{tab:pah_fluxes}. \autoref{fig:pah_integrate} illustrates the flux integration of PAH features.

To estimate uncertainties on the integrated PAH fluxes, we employed a Monte Carlo approach similar to that used by \citet{Bajaj2024AJ....167..127B}. First, we fit a linear baseline to continuum regions adjacent to each PAH feature and computed the standard deviation ($\sigma$) of the residuals between the observed spectrum and the fitted continuum in these regions. We then generated 5000 synthetic spectra by adding gaussian noise with standard deviation $\sigma$ to the observed continuum-subtracted flux. For each realization, we performed numerical integration over the same wavelength intervals used for the PAH features (as described above). The standard deviation of the resulting distribution of integrated flux values was adopted as the 1$\sigma$ uncertainty on the measured PAH flux. This method accounts for both the spectral noise and the sensitivity of the flux integration to local continuum variations.

The flux estimation in continuum subtracted \Spitzer\ spectra (low \& high resolution) follows a similar approach as JWST spectra above. The errors are estimated accounting for flux uncertainties given in CASSISJuice spectra \citep{Arun2025arXiv250614218A}. Due to the weak PAHs in Spitzer SL, we do not decompose the 6.0 and 6.2 \micron\ features and integrate them to get composite flux in the wavelength range of 5.9-6.5. The other PAHs features are integrated in the same regions as the \JWST\ spectra. The fluxes are given in \autoref{tab:pah_fluxes}. 

The continuum subtracted VLT/ISAAC L-band spectrum is integrated between 3.28 and 3.40 \micron to estimate the flux in 3.3 \micron\ PAH feature. The integration of VLT/ISAAC L-band spectrum is shown in \autoref{fig:pah_integrate3_micron}.

\subsection{PAHs in \Spitzer\ \texttt{SL} Spectrum}
The \Spitzer\ observation of \tcha\ in the high resolution mode (AOR - 5641216) taken in 2004 is used previously for various studies \citep{Lahuis2007ApJ...665..492L,Geers2007A&A...476..279G,Brown2007ApJ...664L.107B,Kessler2006ApJ...639..275K}. But the low resolution (\texttt{SL} : R $\sim$ 60) \Spitzer\ spectrum (AOR - 12679424) is not used by previous studies until recently. \cite{Chengyan2025ApJ...978...34X} used the data for establishing the “seesaw effect”, mentioning the data is from \cite{Brown2007ApJ...664L.107B}, but the data reported in the paper is the \texttt{SH+LH} mode. We found significant bumps in the PAH wavelengths in the spectrum, and we performed a continuum subtraction detailed in section 3.2. We report for the first time small but significant detection of PAHs at 6.2, 7.5-9 and 11.2 \micron\ regions in the \Spitzer\ \texttt{SL} spectrum of \tcha\ from 2005.

\begin{figure*}[ht!]
\centering
\includegraphics[width=1.96\columnwidth]{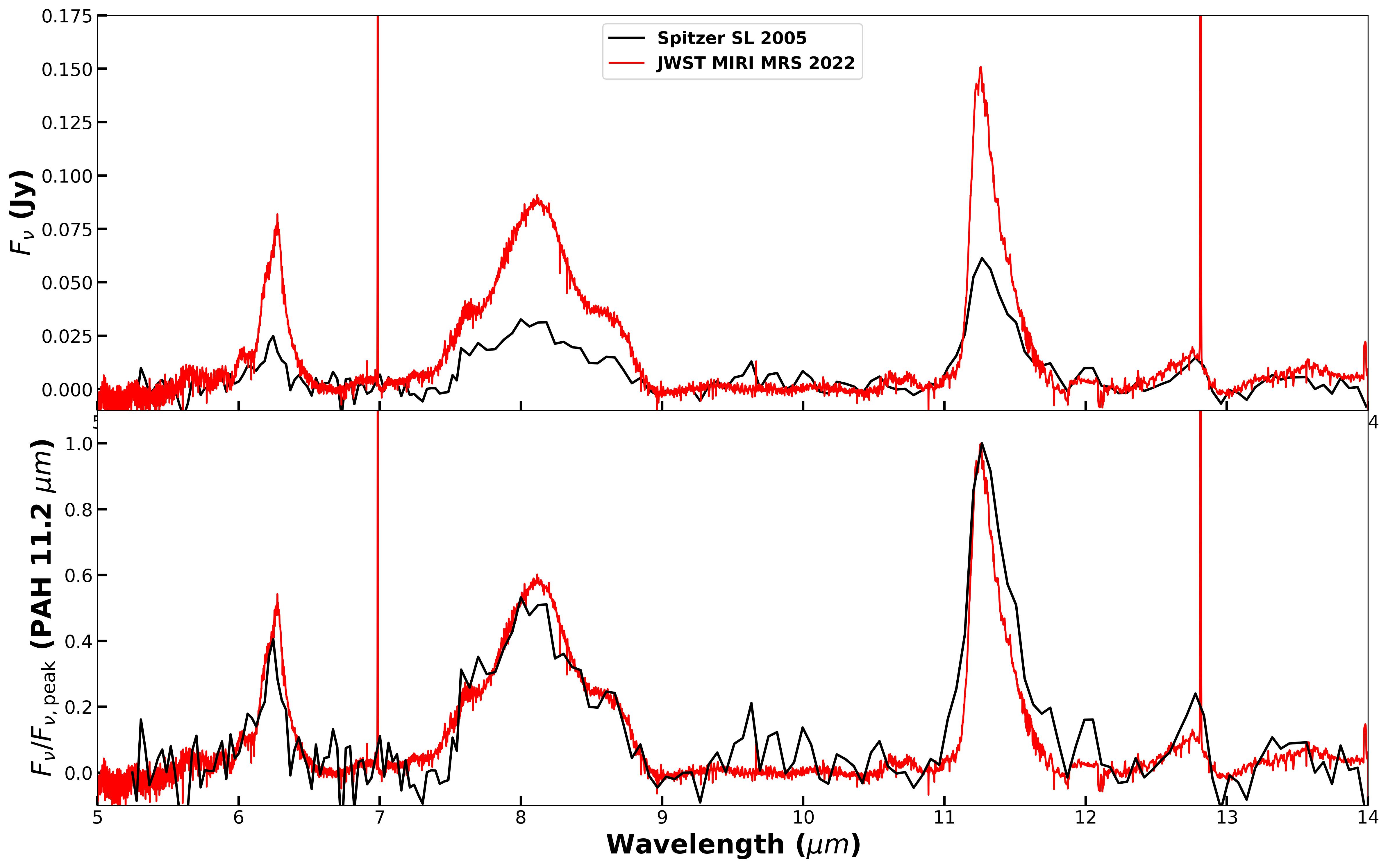}
\caption{Continuum-subtracted mid-IR spectra of \tcha\ from \JWST/MIRI (red) and archival \Spitzer\ \texttt{SL} observations (black). Top panel: Continuum-subtracted spectral comparison showing prominent emission features between 5–14 \micron\ in both \JWST\ and \Spitzer. Bottom panel: Same spectra normalized to the peak of the 11.2 \micron\ PAH feature, emphasizing the ratios have not changed significantly in 18 year timescale.}
\label{fig:pahcompare}
\end{figure*}

The detection of PAHs in the \Spitzer\ \texttt{SL} spectra gives us a unique opportunity to compare the flux ratios. We compared the continuum subtracted \Spitzer\ \texttt{SL} and \JWST\ spectra of \tcha\, which is shown in \autoref{fig:pahcompare}. The bottom panel shows flux of both epochs normalized to the peak of 11.2 \micron\ PAHs. Interestingly, the flux in composite 11.2 \micron\ PAH (11.0 + 11.2) feature has increased $\sim$ 1.75 times from \Spitzer\ \texttt{SL} to \JWST\ epochs, indicative of the higher UV fluorescence due to wall destruction. The normalized figure (bottom panel) shows relative flux of 6.2 \micron\ PAH and 7.7 \micron\ complex with 11.2 \micron\ PAH does not show any significant change in both epochs. The 6.2/11.2 PAH flux ratio is 0.77 $\pm$ 0.17 in the \Spitzer\ epoch and 1.34 $\pm$ 0.04 in the \JWST\ epoch. In the case of 7.7/11.2 ratio, the values are 1.59 $\pm$ 0.17 and 2.9 $\pm$ 0.08, respectively. \cite{Bajaj2024AJ....167..127B} found that the flux ratios at \JWST\ epoch indicate a neutral PAH population in the \tcha\ disk. 
While both ratios nearly double, they stay in the low-ionization PAH regime. In our charge–size analysis (Sect. 4.2), based on the grids of \citet{Maragkoudakis2020MNRAS.494..642M}, these ratios place \tcha\ at $\sim$75\% neutral. For context, highly ionized PAH environments in disks typically exhibit substantially larger $7.7/11.2$ ratios than observed for \tcha. The $7.7/11.2$ ratio 4.2, 6.2, and 11.0 for SR~21A, PDS~144N and, HD~97300, respectively, broadly increasing with stellar UV output \citep{Arun2025arXiv250614218A}.

The 7.7/11.2 ratio has increased slightly in \JWST\ epoch, this indicates to a small increase in ionization \citep{Maragkoudakis2020MNRAS.494..642M}. But the increase does not alter the ionization of the PAH population significantly (Further details are discussed in Section 4.2).

\begin{table}[ht]
\centering
\caption{Measured PAH Fluxes from the mid-IR spectra of \tcha}
\label{tab:pah_fluxes}
\begin{tabular}{ccc}
\hline\hline
& PAH Feature ($\mu$m) & Flux (erg\,cm$^{-2}$\,s$^{-1}$) \\
\hline
VLT/ISAAC & & \\
& 3.3   & $2.242 \times 10^{-12}$ \\
\Spitzer\ & & \\
2004 & 11.0 + 11.2  & $2.85 \pm 0.3 \times 10^{-13}$ \\
     & 12.7\textsuperscript{*} & $6.30 \pm 0.1 \times 10^{-14}$ \\
2005 & $\Sigma$6.2   & $4.17 \pm 0.91 \times 10^{-13}$ \\
     & 7.7   & $8.56 \pm 0.82 \times 10^{-13}$ \\
     & 8.6   & $2.37 \pm 0.62 \times 10^{-13}$ \\
     & 11.0 + 11.2  & $5.40 \pm 0.24 \times 10^{-13}$ \\
     & 12.7\textsuperscript{c}  & $0.41 \pm 0.28 \times 10^{-13}$ \\
\JWST\ & & \\
& $\Sigma$6.2 & $1.27 \pm 0.01 \times 10^{-12}$ \\
& 7.7   & $2.75 \pm 0.004 \times 10^{-12}$ \\
& 8.6   & $4.79 \pm 0.03 \times 10^{-13}$ \\
& 11.0  & $3.34 \pm 0.10 \times 10^{-14}$ \\
& 11.2  & $9.15 \pm 0.02 \times 10^{-13}$ \\
& 12.7\textsuperscript{*}  & $1.37 \pm 0.03 \times 10^{-13}$ \\
\hline
\end{tabular}\\\vspace{0.1cm}
\footnotesize{
*) The 12.7~$\mu$m PAH flux are reported after subtraction of [Ne II] 12.81~$\mu$m line flux.}
\end{table}

\subsection{Variability of 11.2 \micron\ Feature}

The 11.2~\micron\ PAH band is among the most prominent PAH features in astrophysical environments \citep{Tielens2008ARA&A..46..289T,Hony2001A&A...370.1030H}. The profile of PAHs shape primarily traces PAH charge/size and the hardness of the radiation field \citep{Maragkoudakis2020MNRAS.494..642M}, but the band flux mainly tracks the intensity of the UV field. All three spectral epochs for \tcha\ show the 11.2~\micron\ PAH feature. Given the noticeable variability in the continuum, likely due to changes in the inner wall structure, it gives a unique opportunity to investigate the variability in both the flux and profile shape of the 11.2~\micron\ PAH feature in \tcha. 

\begin{figure*}[ht!]
\centering
\includegraphics[width=0.96\columnwidth]{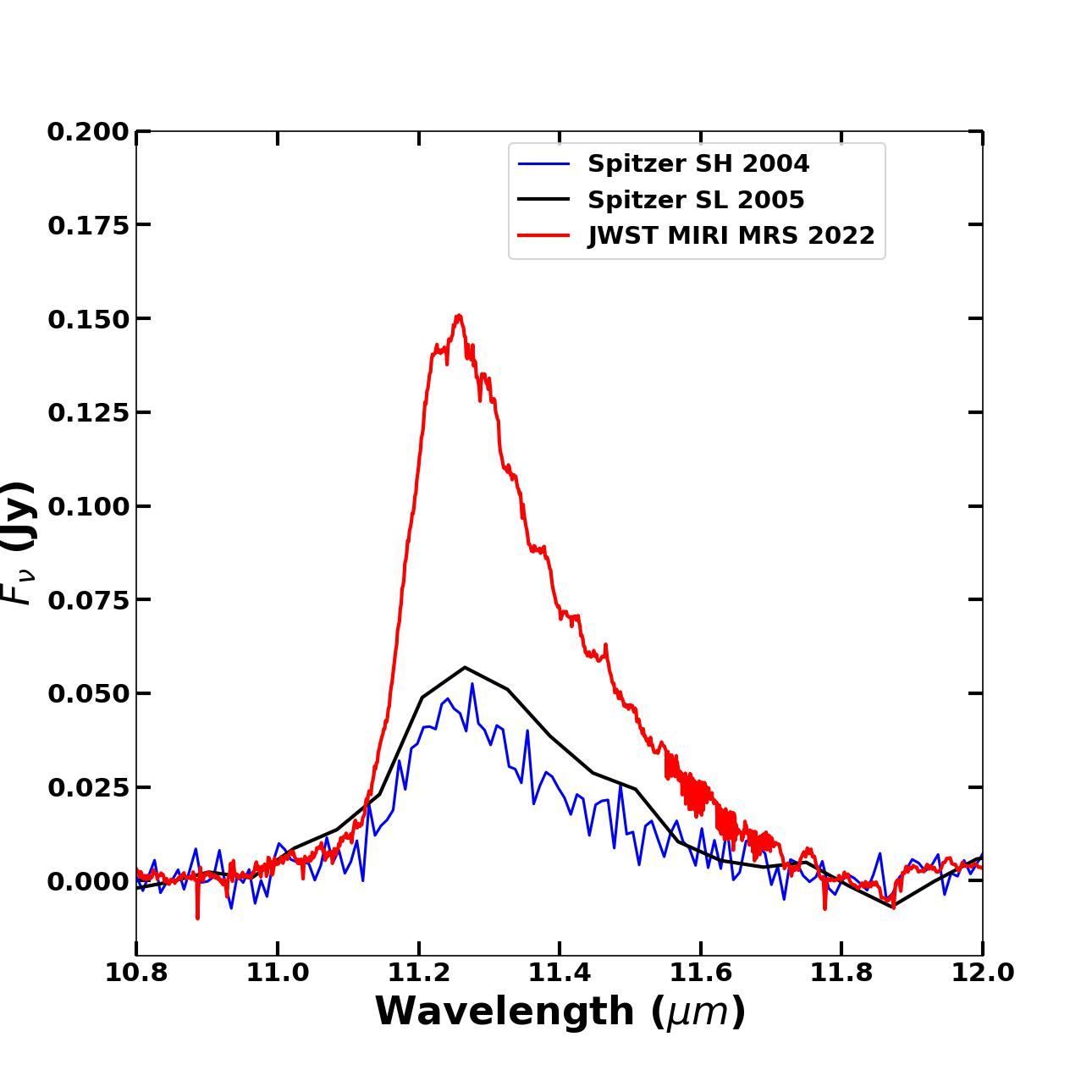}
\includegraphics[width=0.96\columnwidth]{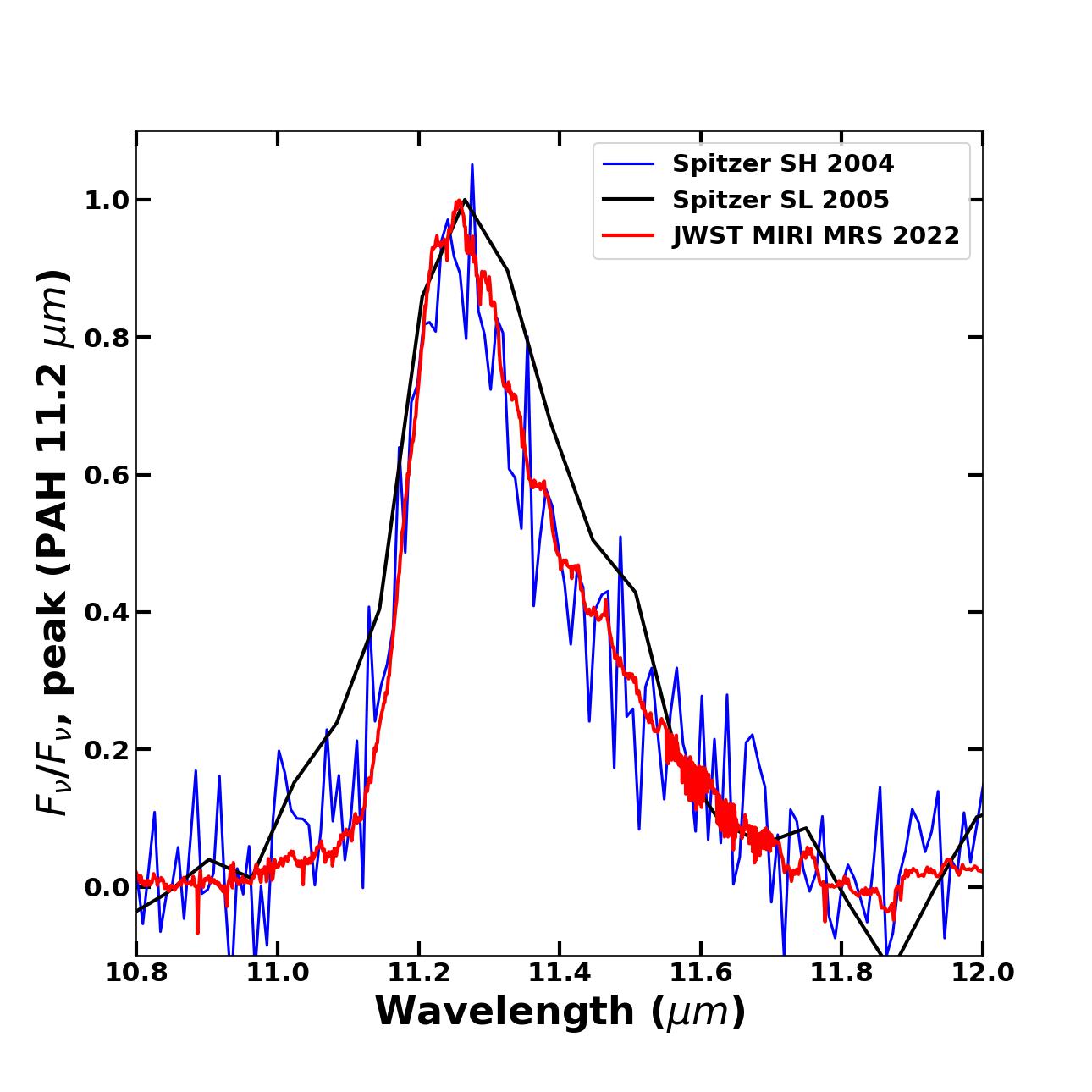}
\caption{
Comparison of the 11.2 \micron\ PAH feature in T~Cha's continuum-subtracted spectra from different epochs (left) and 11.2 \micron\ PAH feature normalized to its peak amplitude(right).:
\textit{Spitzer}~SH (blue}) in 2004, \textit{Spitzer}~SL (black) in 2005, and \textit{JWST}/MIRI (red) in 2022.
(left)~All spectra are normalized to the peak flux of the PAH feature, showing no appreciable change in profile shape.
(right)~The same data plotted in absolute flux units, revealing a factor of $\sim$3 increase in feature strength in the
\textit{JWST} epoch compared to the two \textit{Spitzer} observations
\label{fig:11micron}
\end{figure*}

The two \textit{Spitzer} epochs, which are separated by approximately one year, show small flux variation in the 11.2 \micron\ feature, indicating similar disk structure. However, the \textit{JWST} spectrum (\autoref{fig:11micron} (left)) reveals that the 11.2 \micron\ band flux has increased by a factor of $\sim$3 compared to the \textit{Spitzer} \texttt{SH+LH} measurements in 2004 and by a factor of $\sim$1.75. The estimated flux from \JWST\ and 2005 \Spitzer\ spectra matches with estimates from \cite{Bajaj2024AJ....167..127B}. The three fold increase in the 11.2 \micron\ flux shows that the destruction of the inner disk wall \citep{Chengyan2025ApJ...978...34X} allowed additional UV flux exposure to outer disk surface. This heightened UV exposure excites PAH molecules more efficiently enhancing the flux of the feature \citep{Tielens2008ARA&A..46..289T}.

The investigation of profile morphology also gives a similar conclusion. Despite the flux increase, normalizing the profiles at their peaks indicates that the shape of the 11.2 \micron\ PAH band remains essentially unchanged across the three epochs (\autoref{fig:11micron}). The profile of PAH features is influenced by environmental conditions such as ionization, grain size distribution, and radiation hardness \citep{Maragkoudakis2020MNRAS.494..642M}. The invariance in the 11.2 \micron\ band shape indicates that the underlying PAH population remains largely unaffected by increasing UV intensity. This also, suggests that variations in the UV field intensity (i.e., the number of photons) are not the primary driver of the changes in PAH feature profiles in various astrophysical environments. Instead, the variability may be attributed to the hardness of the UV field (i.e., change in $\lambda$). One caveat here is, among the prominent PAH features, the profiles of 11.2 \micron\ band is considered less sensitive to variations in the radiation field hardness and ionization state. Thus, even if there is a slight increase in hardness of photons due to enhanced accretion episodes in \tcha~\citep{Cahill2019MNRAS.484.4315C}, it is not high enough to affect the profile of 11.2 \micron\ PAH band. Future \JWST\ followups, while the inner wall is persisting, will be able to study the radiation hardness effects on 6.2 and 7.7 \micron\ PAH bands in \tcha\ system. Long-baseline photometry reveals recurrent wall-high/wall-low episodes on month-to-year timescales \citep{Chengyan2025ApJ...978...34X}, suggesting that follow‑ups can be scheduled by monitoring and triggering \JWST\ during sustained ‘high’ states.  

The invariance of the 11.2 \micron\ PAH profile (peak and FWHM) and the PAH ratios relative to 11.2 \micron\ across epochs suggest that the PAH size distribution and ionization state remained largely unchanged despite the factor of $\sim$3 flux increase. The lack of peak shifts or broadening could indicate minimal processing (e.g., fragmentation) of large, neutral PAHs, which dominate the 11.2 \micron\ emission in shielded disk regions \citep{Peeters2002A&A...390.1089P,Bauschlicher2008ApJ...678..316B}. Additionally, the absence of ionization signatures (e.g.,  11.0 \micron\ feature attributed to the CH out-of-plane bending mode of lone hydrogen groups in PAH cations) supports a neutral charge state \citep{Tielens2008ARA&A..46..289T} consistent with the inference from \cite{Bajaj2024AJ....167..127B}.

\subsection{Comparing the PAH Feature with Reference Sources} \label{subsec:compare}

To better understand the newly detected PAH emission features in the \tcha\ system in the \JWST\ spectrum, we compared its continuum-subtracted spectrum with four reference sources that display strong PAH features: Orion \textit{HII} region PDR template from \JWST\ (hereafter ``HII PDR"; \citealp{Chown2024A&A...685A..75C}), the intermediate-mass star HD 97300 (B9) with associated nebulosity \citep{Manoj2011ApJS..193...11M}, the Herbig Ae star PDS 144N (A2) with a edge-on disk with flare morphology \citep{Perrin2006ApJ...645.1272P}, and the T Tauri star SR 21A (G3; \citealp{Geers2006A&A...459..545G}). The \JWST\ spectrum of HII PDR is taken from \cite{Chown2024A&A...685A..75C} released in PDR4ALL\footnote{https://www.pdrs4all.org/} repository and for the remaining young stellar objects, the \Spitzer\ spectra are taken from CASSISJuice. The \Spitzer\ sources are selected for their broad stellar effective temperature (T\textsubscript{eff}) range and availability of \Spitzer\ low and high resolution spectra observations along with strong PAH emission \citep{Arun2025arXiv250614218A}. To compare the PAH features in the \tcha\ spectrum with reference objects, all spectra were first continuum-subtracted using the spline-based method described in Section 3.2. The next part of the section uses the above data to compare PAHs in the spectra in various wavelength ranges.

\begin{figure*}[ht!] 
\centering 
\includegraphics[width=1.9\columnwidth]{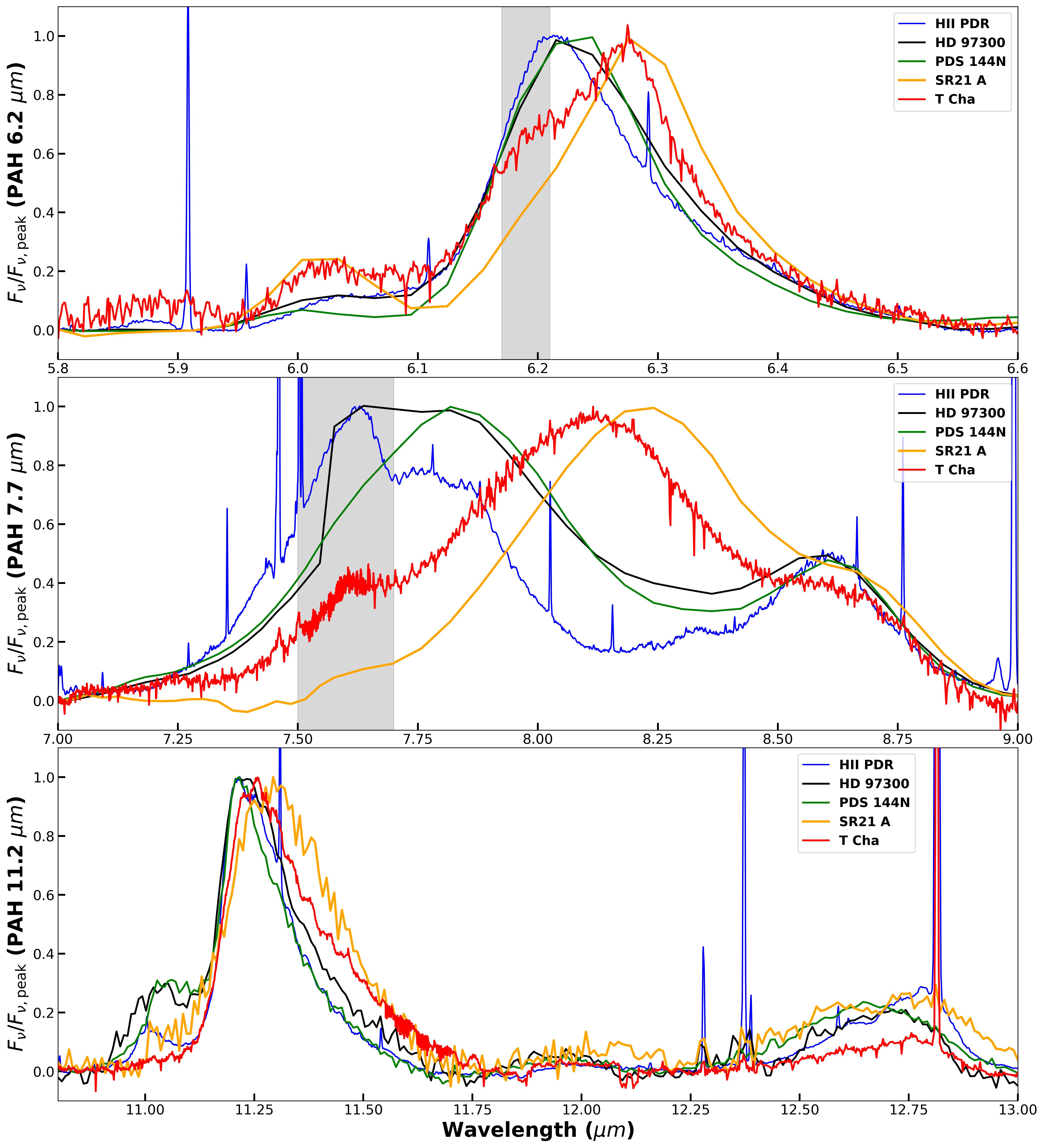}
\caption{ Composite figure showing continuum-subtracted, normalized PAH spectra from \tcha\ and four reference sources. \textbf{Top Panel (6.2 $\mu$m):} Spectra are normalized to unity at their 6.2 $\mu$m peak. Shown are the \JWST\ \textbf{HII PDR} template (blue), \textbf{HD 97300} (black), \textbf{PDS 144N} (green), \textbf{SR 21A} (orange), and \textbf{\tcha} (red). \textbf{Middle panel (7.7 $\mu$m):} The normalized spectra reveal variations in the 7.7 $\mu$m complex, with peak positions shifting from $\sim$7.6 $\mu$m in Class A environments to $\sim$8.1 $\mu$m in \tcha, indicative of different PAH excitation conditions. \textbf{Bottom panel (11.2 $\mu$m):} The 11.2 $\mu$m feature, normalized similarly. The grey band highlights the minor class A feature identified in \tcha.} \label{fig:compare_6p2} 
\end{figure*}

\begin{figure*}[ht!]
\centering
\includegraphics[width=0.95\columnwidth]{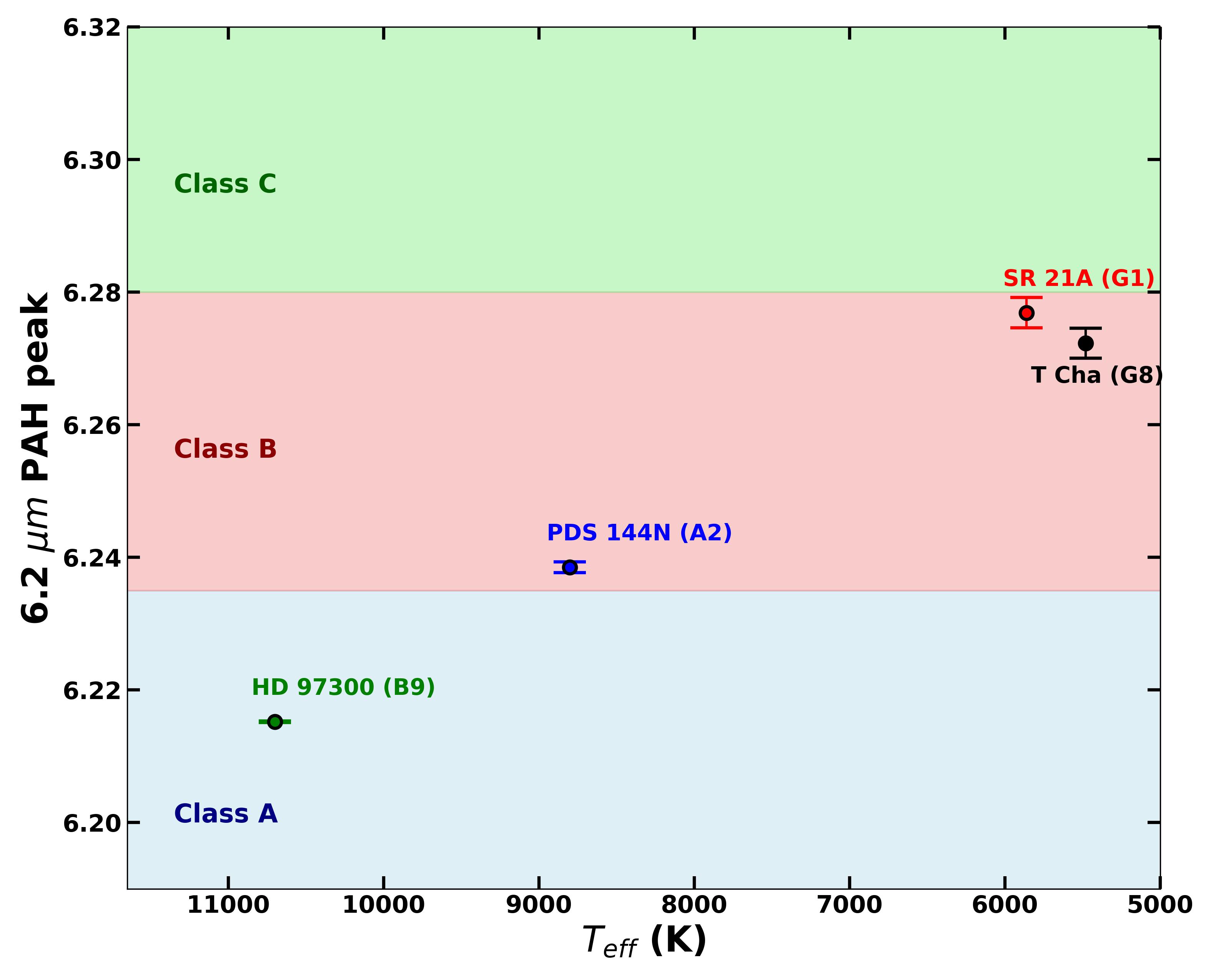}
\includegraphics[width=0.95\columnwidth]{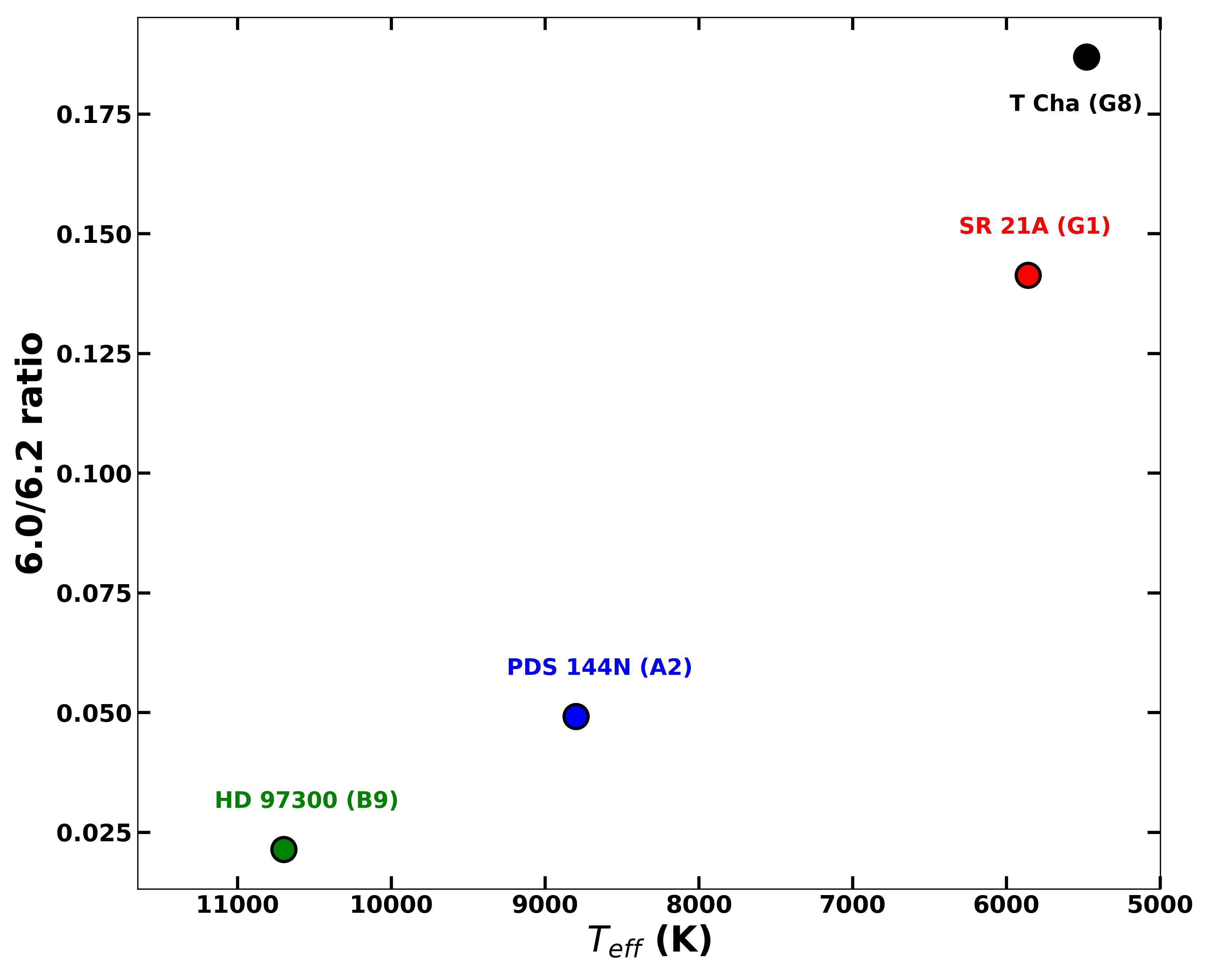}
\caption{Variation of the 6.2 $\mu$m peak position (left) and the 6.0/6.2 (right) ratio with $T_{\text{eff}}$ for \tcha\ and reference stars.}

\label{fig:6_micron_flux}
\end{figure*}
\subsubsection{6.2 \micron\ PAH}

 To compare the 6.2 \micron\ PAH feature in the \tcha\ spectrum with reference objects, we aligned the spectra dividing each by its respective 6.2 \micron\ peak flux, effectively normalizing the peak of 6.2 \micron\ feature to unity (i.e., $F_\nu / F_{\nu,\text{peak}}$). The resulting scaled spectra illustrate shift in peak position, asymmetries in the band wings, and contrast in the overall band shapes. \autoref{fig:compare_6p2} (top panel) shows each spectrum centered on the 5.8--6.6 \micron\ region.

In the section 3.2.2, we found that 6.2 \micron\ feature of \tcha\ (red line) peaks at $\sim6.27$ \micron, shifted toward longer wavelengths relative to most canonical “class A” 6.2 \micron\ as seen in \JWST\ HII PDR template \citep{Peeters2002A&A...390.1089P}. This shift is consistent with border class B/class C PAH emission. Notably, feature profile is broader in its red wing compared to HD 97300 (black line) and PDS144N (green line). The spectra of SR 21A (orange line), being a T Tauri with lower T\textsubscript{eff} shows similar PAH peak with that of \tcha. A major difference in the spectra of \tcha\ is that the component which peaks at 6.19 \micron\ appears to be a weak feature along with the major peak at 6.27 \micron. In all other spectra, the 6.2 \micron\ region looks to be smooth and does not show additional components above the broad feature. Thus, we can say that \tcha\ has a minor class A component in its spectra, which peaks at 6.19 \micron. The class A component is highlighted in \autoref{fig:compare_6p2} (top panel).

Also, we see 6.0 \micron\ weak feature in all sources, which is proposed to be from C=O (carbonyl) stretching vibrations \citep{Peeters2002A&A...390.1089P} or olefinic C=C bonds \citep{Hsia2016ApJ...832..213H}. In the case of lower temperature sources SR21A and \tcha, the ratio of 6.0/6.2 PAH ratio is higher than other two stars and the HII PDR spectra. \autoref{fig:6_micron_flux} shows the variation of 6.2 \micron\ peak and 6.0/6.2 ratio with T\textsubscript{eff}, for \tcha\ and reference sources. The 6.0/6.2 ratio shows a decreasing trend with increasing $T_{\text{eff}}$, indicating that hotter stars tend to exhibit lower values of this ratio suggesting C=O (carbonyl) or olefinic C=C bonds stretching vibrational bonds tend to break in stronger UV fields and are dominant in lower UV fields.

We see a systematic shift in the 6.2 $\mu$m peak position with spectral type further supports a correlation between PAH characteristics and the local stellar radiation environment. SR 21A and \tcha\ aligning within the Class B domain but in the border region between class B and Class C, whereas PDS 144N and HD 97300 has a class B and Class A type PAH features respectively. This also shows that with the decrease of UV hardness, the PAH class shift from C to A \citep{Sloan2007ApJ...664.1144S}. 

Overall, the 6.2 \micron\ PAH feature in \tcha\ emerges as border class B/class C feature linking the system more closely with circumstellar disks that harbor PAHs, rather than with the class A type 6.2 \micron\ band from HII region PDRs or reflection nebulae. It is interesting to note that \JWST\ spectrum of \tcha\ shows a class A pronounced feature peaking around 6.19 \micron\ in the PAH complex.

\subsubsection{7.7 \micron\ PAH Complex}

The 7-9 \micron\ region in intense PAH sources show two major features at 7.7 and 8.6 \micron. We analyze the 7.7 \micron\ PAH band (or “7.7 complex”) of \tcha\ and compare it the reference spectra by dividing the 7.7 \micron\ peak flux on the respective spectra. \autoref{fig:compare_6p2} (middle) presents the resulting scaled profiles for spectra of \tcha\ and the reference sources. Each source shows a different PAH excitation environment, from high-mass stars with reflection nebulosity to lower-mass stars with protoplanetary disks.

For the HII PDR template, the spectrum displays a classical Class A profile peaking near 7.6 \micron\ with a steep-sided structure with a minor broad feature peaking at 7.8 \micron, while HD 97300 shows a transitional “Class AB” profile with comparable maxima at 7.6 and 7.8 \micron. The Herbig Ae star PDS 144N exhibits a profile peaking at approximately 7.8 \micron, consistent with Class B emission, whereas the T Tauri star SR 21A presents a broader, redder feature characteristic of Class C emission. Notably, the spectrum of \tcha\ is shifted to peak near $\sim$8.1 \micron, which is a clear hallmark of a Class C PAH spectrum; however, a minor sub-component at 7.6 \micron\ is also visible, which resembles the class A feature. This presence of class A component in 6.2 and 7.7 \micron\ PAHs are suggestive of a contribution from PAHs populated by a higher UV energy field other than the stellar UV output. The minor class A component is highlighted in \autoref{fig:compare_6p2} (middle panel). The dominant peak and broader red wing around 8.0–8.1 \micron\ overwhelmingly classify \tcha’s 7.7 \micron\ band in the “class~C” regime, parallel to what we find in its 6 \micron\ emission. 

A further point of interest is the relative weakness of the peak of 8.6 \micron\ band with the peak of 7.7 \micron\ band in \tcha. In all the reference sources, the peaks (relative height) of 7.7 relative to 8.6 \micron\ is similar. The 7.7 \micron\ band is attributed to C–C stretching modes, and 8.6 \micron\ is attributed to C–H in-plane bending modes. PAHs with fewer peripheral hydrogen atoms will naturally have a weaker 8.6 \micron\ feature. A possible proposition is the enhancement of the 6.0 \micron\ feature, which is attributed to C=O (carbonyl) groups could lead to a decrease in peripheral hydrogen atoms as it involves the removal of hydrogen atoms from the peripheral positions of the PAH structure. \tcha\ being the lowest mass source, it is possible that higher level of hydrogen substitution might have occurred among the PAH population.

Overall, these observations reinforce the conclusion that \tcha\ hosts a distinctly shifted PAH spectrum consistent with “class C” spectrum. While weak sub component at 7.6 \micron\ match features from a more typical “class A” group suggestive of multiple UV sources with varying hardness illuminating the disk of \tcha.

\subsubsection{11.2 and 12.7 \micron\ PAH Features}

We compared the spectra of \tcha\ with those of our reference sources in the 11–13 \micron\ range after normalizing all profiles to unity at the 11.2 \micron\ peak. In the HII PDR template, HD 97300, and PDS 144N, the 11.2 \micron\ peak is centered at approximately 11.22 \micron, and their overall profiles are nearly identical. In contrast, the 11.2 \micron\ peak from \tcha\ is slightly redshifted to about 11.26 \micron, while the spectrum of SR 21A exhibits the most redshifted peak at approximately 11.29 \micron. These shifts, although present, are much smaller than those observed for the 6.2 and 7.7 \micron\ C–C stretching modes.

Examination of the 11.0 \micron\ feature reveals significant differences among the sources. Both HD 97300 and PDS 144N display a very pronounced 11.0 \micron\ feature relative to the 11.2 \micron\ emission, indicative of a PAH population with a high degree of ionization. The HII PDR spectrum also shows a moderately strong 11.0 \micron\ feature, whereas SR21A exhibits only a weak bump at 11.0 \micron, and \tcha\ shows no discernible 11.0 \micron\ emission, suggesting that its PAHs are predominantly neutral.

Regarding the 12.7 \micron\ feature, the relative \text{fluxes} are comparable among most sources, with the notable exception of \tcha. The 12.7 \micron\ emission is relatively weak in \tcha, implying a lower abundance of duo or trio hydrogen modes within its PAH population \citep{Tielens2008ARA&A..46..289T}. Also, recent study by \cite{Ricca2024ApJ...968..128R} shows that the 11.2 \micron\ PAH is dominated by ``zigzag PAHs" which have high number of solo-hydrogen, and the 12.7 \micron\ PAH are due to arm-chair edge PAHs. This behavior, when considered alongside the differences observed in the 11.0 \micron\ feature, reinforces the conclusion that the PAHs in \tcha\ are mostly neutral and dominated by zigzag PAHs.

\section{Discussion} \label{sec:discussion}

\subsection{Note on Radiation Hardness in \tcha}
The observed line ratios such as [\NeIII]\, 15.55/[\NeII]\, 12.81 $\approx 0.104$ and [\ArIII]\, 8.99/[\ArII]\, 6.98 $\approx 0.019$ in \tcha\ spectrum is low and clearly indicate a soft radiation field in its circumstellar environment \citep{Bajaj2024AJ....167..127B}. These low ratios indicate that very few photons above $\sim$40 eV (required for Ne+++ production) or above $\sim$28 eV (for Ar+++) are present, implying that most of the neon and argon remain singly ionized. Furthermore, the detection of the H$_2$ S(3) line at 9.66\ $\mu$m, together with the non-detection of the commonly observed hydrogen recombination line H\,I (7-6) at 12.37\ $\mu$m \citep[e.g.,][]{Rigliaco2015ApJ...801...31R}, reinforces this conclusion. In such a low-hardness regime, the incident photon energies are predominantly below 13.6\,eV, ensuring that most of the PAH molecules remain neutral and that the UV field does not strongly influence their excitation. This is consistent with the expected spectral output of a G8 T Tauri star, where the radiation is softer compared to that found in high-excitation regions such as the Orion Bar or 30~Dor.

\subsection{Charge and Size of PAH Population}
The model on charge and size of PAHs by \citet{Maragkoudakis2020MNRAS.494..642M} is applied to the PAH flux ratios of \tcha. The model uses emission spectra of PAHs from the NASA Ames PAH IR Spectroscopic Database (PAHdb; \citealp{Mattioda2020ApJS..251...22M}), accounting for a wide range of molecular sizes, structures, and charge states. We use the diagnostic grid in the (11.2+11.0)/7.7 versus (11.2+11.0)/3.3 \text{flux} ratio grid estimated using 6 eV photon energies, enabling simultaneous constraints on PAH ionization and size. The 3.3 $\mu$m PAH flux, is obtained from VLT/ISAAC L-band spectra  reported by \cite{Geers2007A&A...476..279G}. 

\begin{figure}
\centering
\includegraphics[width=\columnwidth]{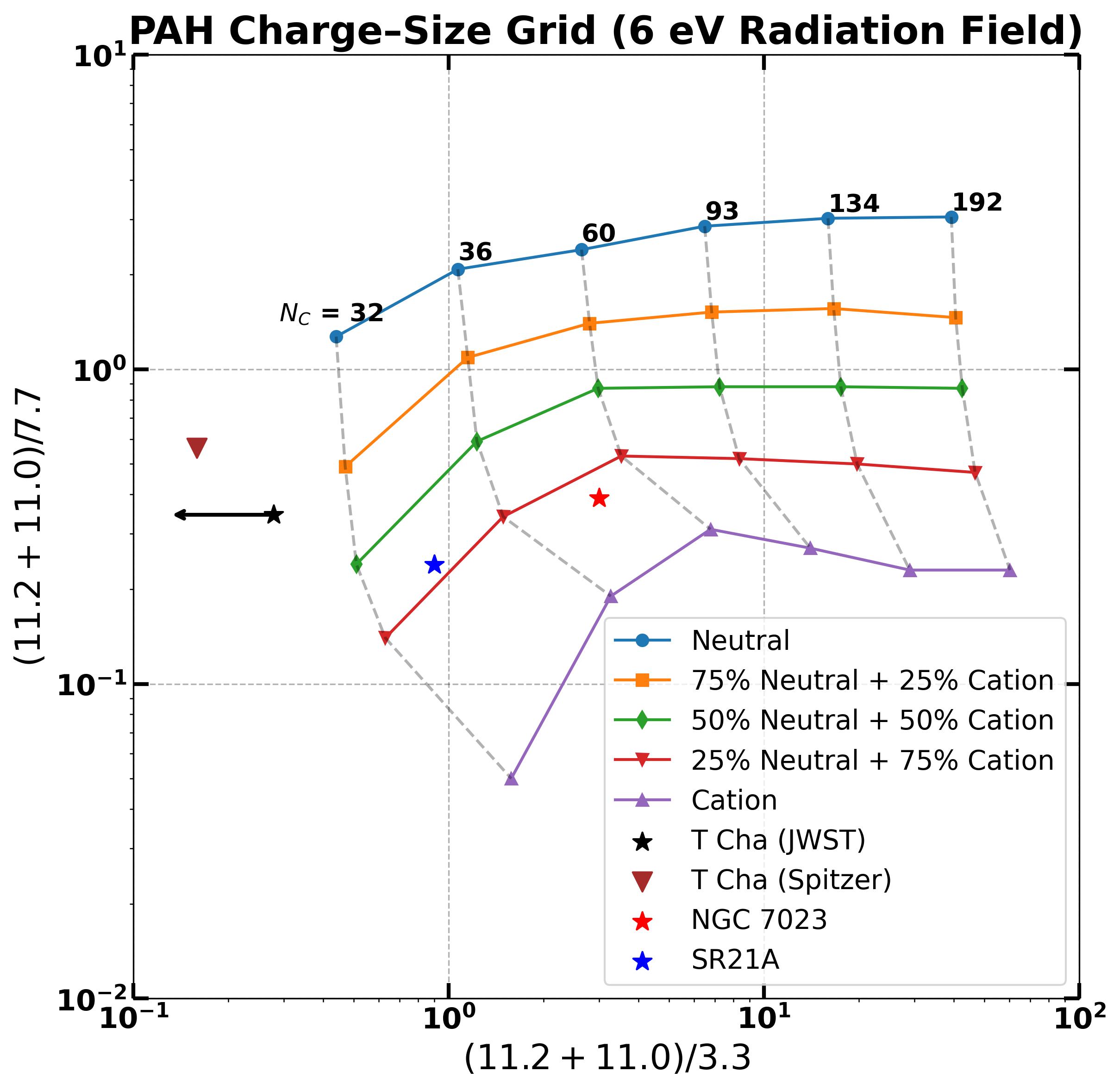}
\caption{Location of \tcha\ observed in \JWST\ (black star) and \Spitzer\ (brown triangle) on the PAH charge–size diagnostic grid constructed using data from \citet{Maragkoudakis2020MNRAS.494..642M}.} The grid shows model tracks for PAH populations with varying neutral-to-cation ratios (colored curves) and dashed grey curves indicate PAH sizes. \tcha\ lies outside the model grid in both epoch, suggesting a low ionization fraction (75\% neutral) and small PAH sizes, with $N_c$ $< 30$ based on extrapolation from the nearest model tracks.

\label{fig:charge_grid}
\end{figure}

The charge-size grid incorporates synthetic spectra for varying neutral-to-cation fractions, allowing application to astrophysical sources with mixed PAH populations. In \autoref{fig:charge_grid}, we present the location of \tcha\ on the PAH charge--size grid generated under a 6 eV radiation field, using PAH fluxes from \JWST\ and \Spitzer. The position of \tcha\ lies distinctly outside the model grid in both ratios, in the lower-left region of the diagram. \cite{Mackie2022} and \cite{Lemmens2023} reported that the 3.3 \micron\ band \text{flux} in modelled PAH spectra used by \cite{Maragkoudakis2020MNRAS.494..642M} is systematically overestimated relative to the CH out-of-plane bending modes in the 10–15 \micron\ range, by approximately 34\%. To ensure a consistent comparison with these diagnostic grids, we scale our measured 11.2/3.3 \micron\ ratios by a factor of 0.66 to account for correct for the overestimation. The location of \tcha\ estimated using \JWST\ and \Spitzer\ in the grid suggests that the average PAH population in \tcha\ is characterized by low ionization (approximately 75\% neutral) and small sizes, corresponding to $N_c$ $< 30$ when extrapolating from the established PAH tracks. The finding of low ionization is consistent with the observation of \cite{Bajaj2024AJ....167..127B}. The ratio estimated with \Spitzer\ \texttt{SL} did not significantly move in the grid, suggestive for a very small change of ionization and size. One important point to consider is that the 3.3 \micron\ PAH flux estimated from the L band spectra is taken on 18-04-2006, which is used in both \JWST\ and \Spitzer\ ratios. The timeline of data matches more with \Spitzer\ epoch and is not at the wall destruction phase. So we assume that 3.3 $\mu$m should increase in the \JWST\ epoch similar to 11.2 \micron\ PAHs, the data point will move leftward horizontally, making this a upper limit in the x-axis. The figure also shows NGC 7023 a reflection nebulae and the reference star SR21A. The NGC 7023 is illuminated by a B2 type star, and shows a higher number of $N_c$ and highly ionized \citep{Maragkoudakis2020MNRAS.494..642M}. The G2 type star SR21A shows higher ionization and $N_C$ than \tcha.

\subsection{Origin of the Weak Class A PAH Sub-component} 

The PAH spectrum of \tcha\ is dominated by a class C profile; however, detailed inspection of the 6.2 and 7.7 \micron\ features reveals the presence of a weak sub-component that peaks at approximately 6.19 \micron\ and 7.6 \micron, respectively. We propose that this sub-component arises from a localized enhancement in the high-energy UV flux produced by disk accretion.

Transitional disks such as that of \tcha\ are characterized by a large inner dust gap \citep[e.g.,][]{Espilliat2014prpl.conf..497E}. The reduction or removal of the inner dust wall not only exposes the outer disk to more stellar radiation but also permits UV photons from accretion hot spots to reach the disk surface. Magnetospheric accretion models predict that matter from the disk is funneled along magnetic field lines, forming shock-heated hot spots on the stellar surface with temperatures that can exceed 10,000 K \citep[e.g.,][]{Calvet1998ApJ...509..802C}. Although the underlying G8 stellar photosphere emits a relatively soft UV field that produces the predominant class C PAH emission, the localized hard UV excess from these hot spots can increase the ionization fraction of nearby PAH molecules. 

The UVES observations by \citet{Cahill2019MNRAS.484.4315C} reveal that the mass accretion rate in \tcha\ varies by more than an order of magnitude, from $\log(\dot{M}_{\rm acc}) \approx -8.8$ to $-7.5$~$M_\odot~{\rm yr}^{-1}$. Such a significant increase in accretion rate would naturally lead to an enhancement in the accretion-driven UV excess hotspots, particularly in the far-ultraviolet (FUV) regime. This elevated UV flux could locally harden the radiation field at the disk surface, especially while the inner wall is absent \citep{Chengyan2025ApJ...978...34X}. We propose that during these high-accretion phases, the resulting burst of hard UV photons from the magnetospheric accretion hotspots is capable of transiently altering the ionization state of PAHs near the inner rim of the outer disk. This process could give rise to weak but detectable sub-components in the PAH spectrum, including features peaking at $\sim$6.19~$\mu$m and 7.6~$\mu$m, characteristic of class A profiles. The high resolution and sensitivity of \JWST, enabled the detection of these sub-components in the two PAH features from the circumstellar disk for the first time. More such detections can enhance our knowledge on PAH excitation near multiple UV sources with varying hardness. To test this scenario, further observations could be directed at correlating accretion variability (tracked via UV excess or emission line diagnostics) with changes in the PAH feature ratios. \tcha\ presents a compelling science case for conducting a multi-epoch near- and mid-IR study.

\section{Summary and Outlook} \label{sec:conclusions}

This study presents a detailed analysis of PAH emission in transitional disk around the G8 T Tauri star, \tcha\ using archival \Spitzer\ data and recent \JWST/MIRI MRS observations separated by nearly two decades. The disk has undergone a significant evolution in the structure, particularly the destruction of the inner dust wall, which revealed the PAH emission features which was not previously reported. 

\begin{itemize}
    \item The continuum subtracted \Spitzer\ \texttt{SL} data of \tcha\ observed in 2005 showed 6.2, 7.7, and 8.6 \micron\ PAH features, which was not reported before.
    \item The relative \text{fluxes} of 6.2 \micron\ PAH to 11.2 \micron\ PAH is 0.77 $\pm$ 0.17 in the \Spitzer\ epoch and 1.36 $\pm$ 0.02 in \JWST\ epoch, both values corresponds to low ionization (approximately 75\% neutral).
    \item The 11.2 \micron~ PAH band of \tcha\ has increased in flux by a factor of ~3 between the \Spitzer\ and \JWST\ epochs spanning 18 years but shows no change in profile shape or peak position, also indicating a stable, neutral PAH population despite UV field variability.
    \item PAH band fluxes at 6.2, 7.7, 8.6, and 12.7 \micron\ have increased at least twofold in \JWST\ epoch, resulting from increased UV excitation after inner-wall destruction.
    \item The 6.2 and 7.7 \micron\ PAH bands show redshifted peaks and broad profiles consistent with class C spectra, commonly associated with low-UV environments and partially hydrogenated or aliphatic-rich PAHs.
    \item Weak class A-like components at 6.19 and 7.6 \micron\ suggest the presence of localized harder UV fields, possibly from accretion hotspots, contributing to a mixed PAH population.
    \item Enhanced 6.0/6.2 PAH ratio of \tcha\ indicates higher carbonyl (C==O) substitution or increased olefinic C==C bonds than higher mass sources.
    \item The weaker 12.7/11.2 PAH ratio shows that lower abundances of duo- or trio-hydrogen modes in the carbon chain suggesting ‘zigzag’ molecular structure for PAHs in the disk of \tcha. 
    \item Using PAH diagnostic grids, the PAH population in \tcha\ is found to be predominantly neutral with small carbon skeletons ($N_c <$ 30), consistent with a soft UV radiation field from a G8-type star.

\end{itemize}

\tcha\ emerges as a benchmark system where the persistence of class C PAH emission with a class A sub component with the pronounced variability in its inner disk continuum. Coordinated, triggerable multi‑epoch \JWST\ observations based on photometric monitoring can directly test how radiation hardness and PAH charge respond across wall‑high and wall‑low phases and can help understand changes in PAHs with accretion and disk variability.

\begin{acknowledgments}
Thanks to the referee for providing insightful comments and suggestions that significantly improved this paper. Author thanks Mr. Naman Bajaj providing the \JWST\ spectrum of \tcha. The JWST data used in this paper can be found in MAST: \dataset[10.
17909/dhmh-fx64]{https://doi.org/10.17909/dhmh-fx64}. Thanks to Dr. Vincent Geers for providing VLT/ISAAC spectra. Also, thanks to Dr. Shridharan Baskaran and Dr. Akhil Krishna R for their valuable comments on this work and wonderful discussions. The author thanks Dr. Merlin Thomas for the encouragement during this study. This work made use of Astropy:\footnote{http://www.astropy.org} a community-developed core Python package and an ecosystem of tools and resources for astronomy \citep{astropy:2013, astropy:2018, astropy:2022}. Also, ChatGPT (OpenAI 2023) is used for assistance in correcting typos and grammar, though full responsibility for the manuscript’s content remains our own.
 
\end{acknowledgments}

\begin{contribution}

The sole author was responsible for the conceptualization, data analysis, interpretation, software development, visualization, and writing of the manuscript.

\end{contribution}

%
\facilities{\JWST (MIRI), \Spitzer (IRS), VLT/ISAAC} 

\software{astropy \citep{astropy:2013}
          }


\appendix

\section{Initial Gaussian parameters of 6.2 \micron\ PAH band}

\autoref{tab:6um_init_params} gives the initial gaussian parameters. The FWHM values are converted to standard deviations for Levenberg--Marquardt fitter and bounds of $\pm0.0025\,\mu$m is applied. The bound was set on mean values at $\pm0.025\,\mu$m.
\setcounter{table}{0}		
\renewcommand{\thetable}{A\arabic{table}}
\begin{table}[h]
\centering
\caption{Initial Gaussian Parameters for the 6 \micron\ Region in T~Cha}
\label{tab:6um_init_params}
\begin{tabular}{cccc}
\hline\hline
Component & Center        & FWHM         & Amplitude \\
          & ($\mu$m)      & ($\mu$m)     & ($\mathrm{erg\,s^{-1}\,cm^{-2}\,\text{\AA}^{-1}}$)\\
\hline
C1 & 6.01  & 0.067 & 1.5$\times10^{-16}$ \\
C2 & 6.075 & 0.080 & 1.5$\times10^{-16}$ \\
C3 & 6.20  & 0.080 & 4.0$\times10^{-16}$ \\
C4 & 6.29  & 0.095 & 6.0$\times10^{-16}$ \\
C5 & 6.39  & 0.135 & 1.5$\times10^{-16}$ \\
C6 & 6.50  & 0.062 & 1.5$\times10^{-16}$ \\
\hline 
\end{tabular}
\end{table}

\section{Integrating PAH \text{fluxes}}

\setcounter{figure}{0}          
\renewcommand{\thefigure}{B\arabic{figure}}  
\begin{figure*}[h] 
\centering 
\includegraphics[width=1\columnwidth]{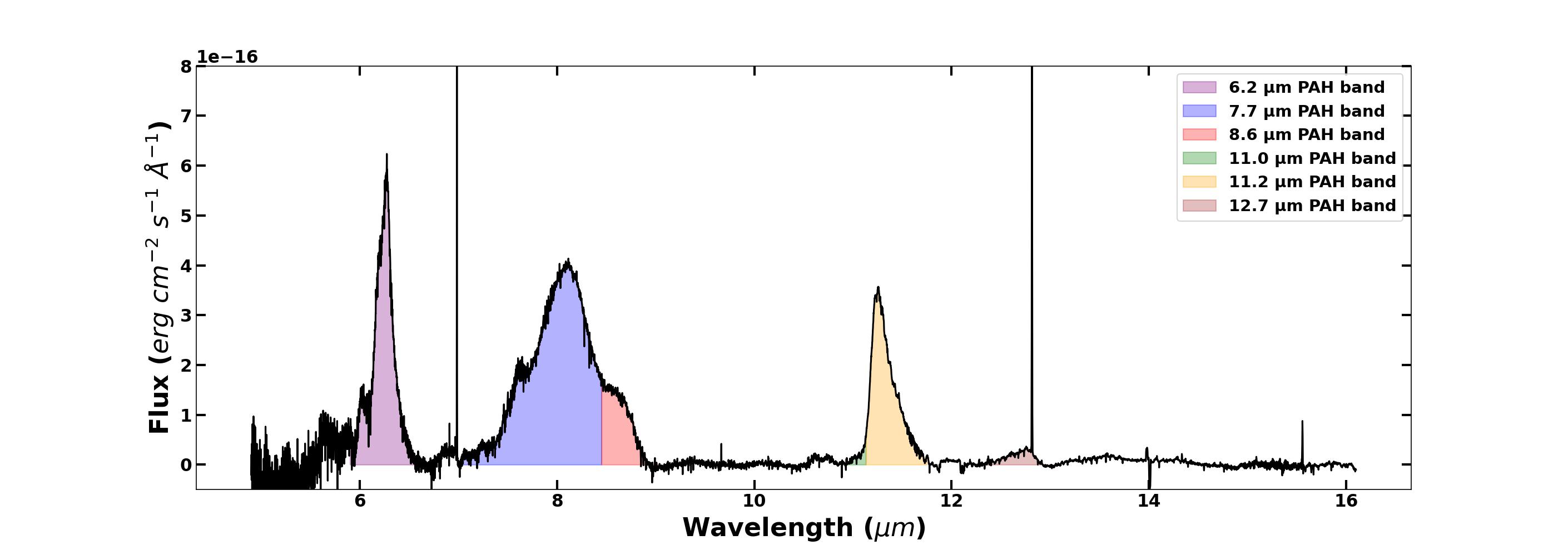}
\includegraphics[width=1\columnwidth]{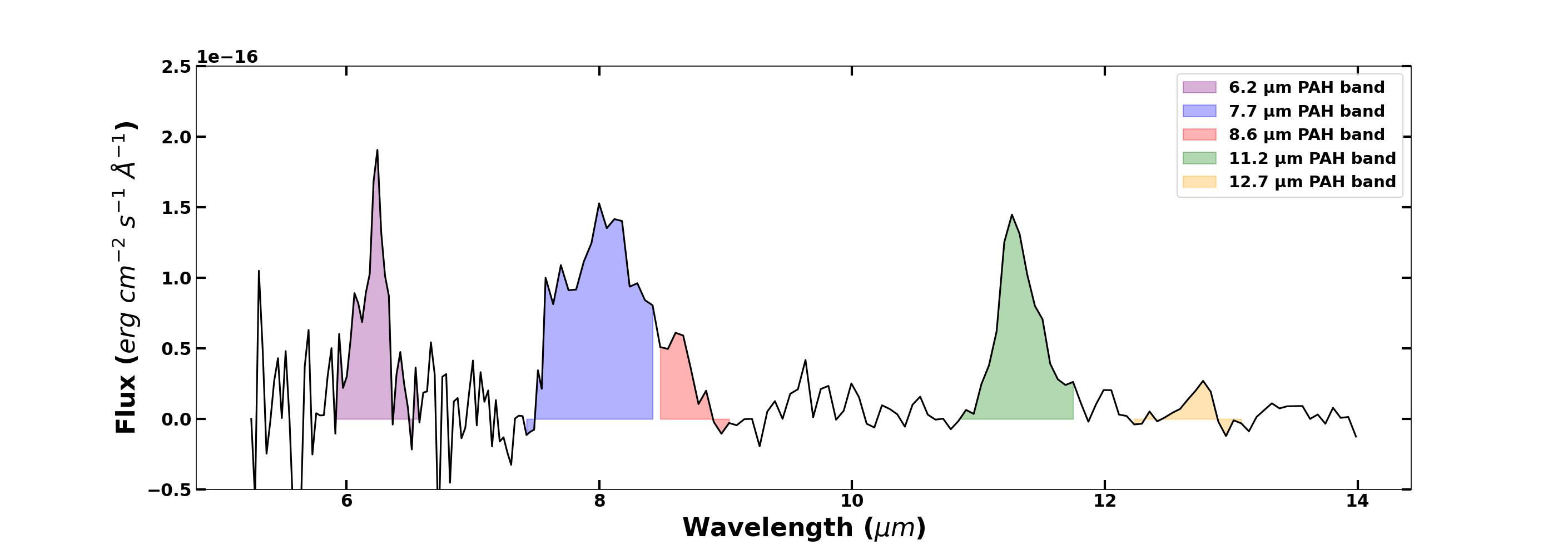}

\caption{Continuum-subtracted \JWST/MIRI (Top panel) and \Spitzer\ (Bottom panel) spectra of \tcha\ highlighted with integrated PAH regions. The final flux at 12.7 \micron\ PAH band is estimated after the subtraction of fitted [NeII] flux in the case of \JWST\ spectrum.} \label{fig:pah_integrate} 
\end{figure*}
\begin{figure*}[ht!] 
\centering 
\includegraphics[width=1\columnwidth]{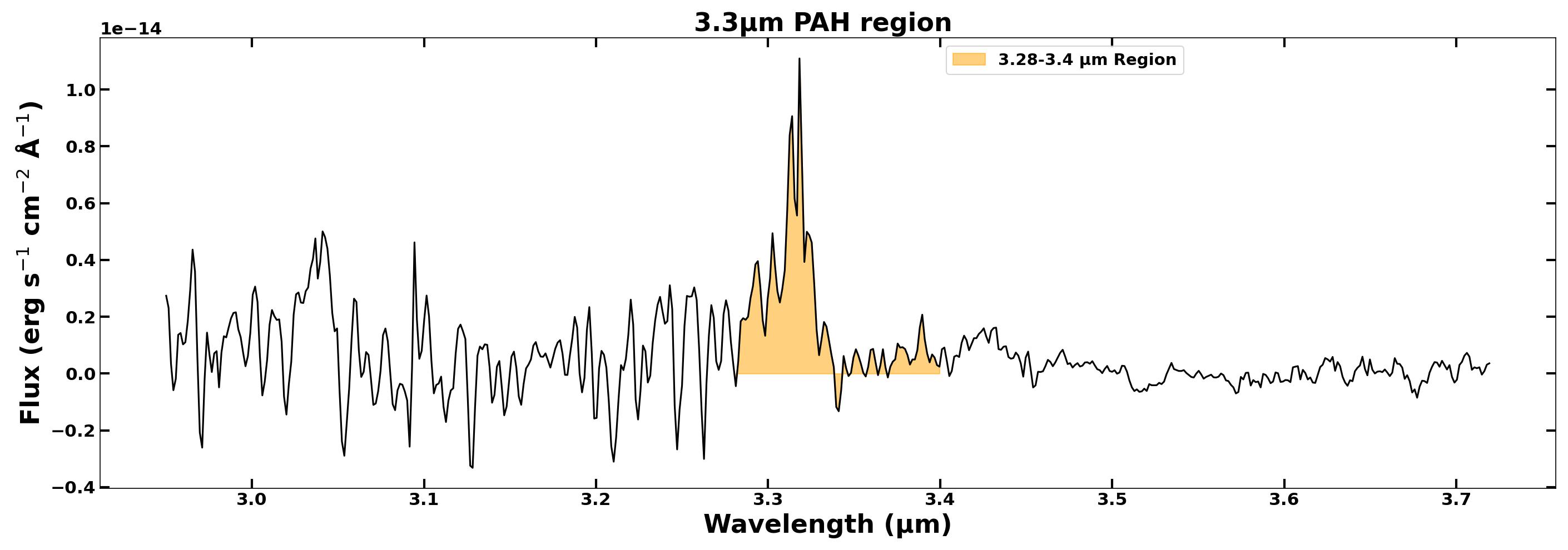}

\caption{Continuum-subtracted ISAAC L band spectrum of \tcha\ highlighting 3.3 \micron\ integrated PAH regions taken from \cite{Geers2007A&A...476..279G}.} \label{fig:pah_integrate3_micron} 
\end{figure*}

\bibliography{sample7}{}
\bibliographystyle{aasjournal}



\end{document}